\documentstyle[preprint,aps,eqsecnum,epsfig]{revtex}
\tightenlines
\def\pmb#1{\setbox0=\hbox{#1}%
     \kern-.025em\copy0\kern-\wd0
      \kern.05em\copy0\kern-\wd0
       \kern-.025em\raise.0433em\box0}

\def\beq{\begin{equation}}
\def\eeq{\end{equation}}
\def\bea{\begin{eqnarray}}
\def\eea{\end{eqnarray}}
\begin{document}
\title{\bf Octupole response and stability of spherical shape in heavy 
nuclei.}
\author{ V. I. Abrosimov$^{\rm a}$, O.I. Davidovskaya$^{a}$,
A. Dellafiore $^{\rm b}$, F. Matera$^{\rm b}$\\
\it $^{\rm a}$ Institute for Nuclear Research, 03028 Kiev, Ukraine\\
$^{\rm b}$ \it Istituto Nazionale di Fisica Nucleare and Dipartimento
di Fisica,\\
\it Universita' di Firenze, via Sansone 1, 50019 Sesto F.no (Firenze), Italy}
\maketitle
\begin{abstract}
The isoscalar octupole response of a heavy spherical nucleus is 
analyzed in a semiclassical model based on the linearized Vlasov 
equation. The octupole strength function is evaluated
with different degrees of approximation. The zero-order fixed-surface 
response displays a remarkable concentration of strength in the 
$1\hbar\omega$ and $3\hbar\omega$ regions, in excellent agreement 
with the quantum single-particle response. The collective 
fixed-surface response reproduces both the high- and low-energy 
octupole rsonances, but not the low-lying $3^{-}$ collective states,
while the moving-surface response function gives a 
good qualitative description of all the main features of the octupole 
response in heavy nuclei. The role of triangular 
nucleon orbits, that have been related to a possible instability of 
the spherical shape with respect to octupole-type deformations, is 
discussed within this model. It is found that, rather than creating 
instability, the triangular trajectories are the only classical orbits 
contributing to the damping of low-energy octupole excitations.
\end{abstract}
\vspace{.5 cm}
PACS: 24.10.Cn, 24.30.Cz
\vspace{.5 cm}

Keywords: Vlasov equation, isoscalar octupole resonances, octupole 
surface modes

\section{INTRODUCTION}
It is well known that there is an intimate connection between the 
shell structure in quantum systems like nuclei and metallic clusters and the 
properties of classical trajectories within these systems (see e.g. \cite{b&m},
p. 579). In particular, for nuclei it has been argued that closed orbits of 
triangular shape might lead to an instability of the spherical shape 
against octupole-type deformations in the region beyond $^{208}$Pb 
\cite{b&m}, p.560. Here we would like to investigate in 
detail this 
possibility by using a semiclassical theory of nuclear response based 
on the linearized Vlasov equation \cite{bri,ads}. Instabilities are 
expected to show up as some kind of pathological behaviour in 
response functions (vanishing eigenfrequencies and
diverging response) and the semiclassical theory 
of \cite{bri} and \cite{ads}, that has already been shown to give good 
qualitative results for lower multipolarities \cite{adm1,adm2,adm3},
is an ideal tool to 
study the role played by classical trajectories in determining the response 
of large quantum systems. Of course quantum corrections are expected 
to modify the results of this theory \cite{koh,dmb}, especially at very 
low energy, however having a 
clear picture of what should be expected already at the classical 
level might help in making progress.
We study the isoscalar octupole response of a sample ``nucleus'' 
of $A=208$ nucleons contained in a square-well potential of radius 
$R=r_{0} A^{\frac{1}{3}}$, with $r_{0}=1.2\,{\rm fm}$. This is done 
with different degrees of approximation: in a first  
approximation, discussed in Sect. II, the interaction between nucleons 
is neglected, while in Sect. III this interaction is taken into 
account in a separable approximation. Finally in Sect. IV the effect 
of surface vibrations on the octupole response function is included.

Because of an interesting scaling 
property of the zero-order semiclassical response, the $A$-dependence  can be 
factorized and becomes trivial. Thus, although we perform explicit 
calculations only for $A=208$, our results on stability apply
also to the region beyond $^{208}$Pb.

\section{Single-particle response (fixed surface)}

Following Ref. \cite{b&m}, the connection between shell structure and 
classical trajectories is most easily illustrated by considering the 
quantum energy levels of a particle in a spherical potential (no 
spin-orbit interaction). In this 
case the energy levels depend on the radial and angular momentum 
quantum numbers  $n_{r}$ and $l$ only. For sufficiently large 
values of these quantum numbers, the energy difference between 
two neighbouring levels can be approximated as
\beq
\label{edi}
\epsilon(n_{r}',l')-\epsilon(n_{r},l)\approx
(n_{r}'-n_{r})\frac{\partial \epsilon}{\partial n_{r}} +
(l'-l)\frac{\partial \epsilon}{\partial l}\,.
\eeq

Now the important point is that the derivatives
${\partial \epsilon}/{\partial n_{r}}$ and
${\partial \epsilon}/{\partial l}$ are essentially the derivatives 
of the classical Hamiltonian with respect to the corresponding action 
variable, hence they can be recognized as the radial $(\omega_{r})$ 
and angular $(\omega_{\varphi})$ frequencies 
of the classical orbits respectively. Thus
\beq
\label{edi2}
\epsilon(n_{r}',l')-\epsilon(n_{r},l)\approx(n_{r}'-n_{r}) \omega_{r}+
(l'-l) \omega_{\varphi}\,,
\eeq 
(we use units such that $\hbar=c=1$).
This combination of frequencies is exactly that appearing in the 
denominator of the zero-order response function obtained in \cite{bri} from 
the solution of the linearized Vlasov equation with fixed-surface 
boundary conditions. Moreover, from a comparison of the zero-order 
Vlasov response function with the analogous quantum propagator 
\cite{dmb}, it 
can be seen that in the Vlasov propagator the quantum matrix 
elements of the excitation operator are replaced by appropriate 
Fourier coefficients that can be evaluated as integrals along the 
classical trajectories. 
Since here we are interested in the octupole response, we report the 
propagator describing the response of a nucleus to an 
external field of the type
\beq
\label{exf}
Q({\bf r})=r^{3}Y_{3M}({\bf \hat r})
\eeq
in the single-particle approximation. This propagator is given by 
\cite{adm2}
\beq
\label{fsr}
{\cal R}_{L=3}^{0}(s)=
\frac{9A}{8\pi}\frac{R^{6}}{\epsilon_{F}}\sum_{n=-\infty}^{+\infty}
\sum_{N=\pm 1,\pm 3}C^{2}_{3N} \int_{0}^{1} dx x^{2}
~s_{nN}(x){{{\Big (}Q^{(3)}_{nN}(x)/R^{3}{\Big )^{2}}\over {s+i\varepsilon-s_{nN}(x)}}}.
\eeq

Instead of the frequency $\omega$, as independent variable we have used 
the dimensionless quantity $s=\omega/(v_{F}/R)$, as a consequence the
$A$-dependence of this propagator is factorized as $AR^{6}\propto A^{3}$. The 
eigenfrequencies (\ref{edi2}) are accordingly replaced by the 
functions (for a square-well potential)
\beq
s_{nN}(x)=\frac{n\pi+N\alpha(x)}{x}.
\eeq

The variable $x$ is related to the classical angular momentum 
$\lambda$ of a nucleon. The relation is $x=\sin\alpha$,
where $\alpha$ is the angle spanned by the radial vector when the 
particle moves from the inner to the outer turning point. Clearly, for 
a square-well potential one has
$\cos\alpha=\lambda/\bar{\lambda}$, where 
$\bar{\lambda}$ is the maximum particle angular momentum $\bar{\lambda}=p_{F}R$,
$p_{F}$ is the Fermi momentum, 
while $v_{F}$ and $\epsilon_{F}$ are the corresponding velocity and energy.
The sum over discrete angular momentum values of the quantum 
propagator is replaced here by the integration over $x$, thus 
the particle angular momentum is treated as a continuous variable.

The quantities $C_{3N}$ in Eq. (\ref{fsr}) are classical limits of the 
Clebsh-Gordan coefficients coming from the angular integration of the 
quantum matrix elements \cite{dmb}. Their explicit value is $C_{3 
\pm1}=\frac{\sqrt{3}}{4}$ and $C_{3 \pm3}=\frac{\sqrt{5}}{4}$. In 
principle the integer $N$ takes values between $-L$ and $L$, however 
only the coefficients $C_{LN}$ where $N$ has the same parity as $L$ 
are nonvanishing.
The coefficients $Q^{(3)}_{nN}(x)$ appearing in the numerator of Eq. 
(\ref{fsr}) have been defined in Ref. \cite{bri}, they are essentially the classical
limit of the radial matrix 
elements of the octupole operator $r^{3}$
and can be evaluated explicitly as
\beq
\label{fou3}
Q^{(3)}_{nN}(x)=(-)^{n}R^{3}\frac{3}{s^{2}_{nN}(x)}{\Big(}1+\frac{4}{3}N
\frac{\sqrt{1-x^{2}}}{s_{nN}(x)}
-\frac{2}{s^{2}_{nN}(x)}+4(|N|-1)\frac{1-x^{2}}{s^{2}_{nN}(x)}{\Big)}\,.
\eeq
For $N=\pm 1$ this expression coincides with that appearing in the 
(compression) dipole response (cf. Eq.(A.5) of Ref. \cite {adm2}),
however for the octupole response we also need terms with $N=\pm 3$.
These new modes have an interesting property since 
the associated eigenfrequencies $s_{nN}(x)$ can vanish in the interval 
$0<\alpha(x)<\frac{\pi}{2}$ [the equation $(n\pi + N\alpha)=0$ has a 
solution for $\alpha=\frac{\pi}{3}$, corresponding to closed triangular 
orbits]. In Ref.\cite{b&m} it has been pointed out that the vanishing 
of this eigenfrequency might give rise to a possible instability 
against octupole-type deformations in nuclei heavier than $^{208}$Pb.
Although at first sight it might seem that the coefficients 
(\ref{fou3}) would diverge when $s_{nN}(x)\to 0$, it is actually 
possible to check that
\beq
\label{lim}
\lim_{x\to\frac{\sqrt{3}}{2}}Q^{(3)}_{\mp1\,\pm3}(x)=-\frac{1}{4}R^{3}\,.
\eeq

The very fact that this limit is finite is important for our 
discussion about the role of triangular nucleon orbits.
The linear response function in the single-particle approximation 
is well-behaved because
\beq
\label{imro}
\lim_{s\to 0}{\Big (}{\rm Im}\, {\cal R}_{3}^{0}(s){\Big )}=0\,,
\eeq
and
\beq
\label{rero}
\lim_{s\to 0}{\Big (}{\rm Re }\,{\cal R}_{3}^{0}(s){\Big 
)}=-\frac{r_{0}^{6}}{8\pi\epsilon_{F}}A^{3} \,,
\eeq
a result that we shall need later.
Since the zero-order octupole strength function is proportional 
to the imaginary part of the response function (\ref{fsr}), we can see 
that the contribution of triangular trajectories to the octupole 
strength function  does not give any pathology, at 
least at the single-particle level. Of course instabilities could 
still arise in the collective response, so we study also this 
response by using the same model, however, before doing this we look 
more in detail to the single-particle octupole strength given by the 
Vlasov theory.

The zero-order octupole strength function 
$S_{L=3}(E)$, given by ($E=\hbar\omega$)
\beq
S_{L=3}(E)=-\frac{1}{\pi}{\rm Im{\cal R}_{3}^{0}(E)}\,,
\eeq
is shown in Fig.1.
It can be seen that the single-particle octupole strength is concentrated
in two regions around 8 and 24 MeV. As pointed out already in \cite{bri}, in this 
respect our semiclassical response is strikingly similar to the quantum 
response, which is concentrated in the $1\hbar\omega$ and 
$3\hbar\omega$ regions.  The modes that contribute most are those 
with $(n,N)=(0,1)\,,(1,-1),\,(-1,3),\,(2,-3),$ for the $1\hbar\omega$ 
region and with $(n,N)=(0,3)\,,(1,1),\,(2,-1),\,(3,-3),$ for the 
$3\hbar\omega$ region. This concentration of strength is quite remarkable 
because our static distribution, which is taken to be of the 
Thomas-Fermi type $f_{0}({\bf r}, {\bf 
p})\propto \theta(\epsilon_{F}-h_{0}({\bf r}, {\bf p}))$,
does not include any shell effect, however, because of the close 
connection between shell structure and classical trajectories 
expressed by Eqs. (\ref{edi2}) and (\ref{fou3}), we still obtain a 
stregth distribution that is very similar to the one  
usually attributed to shell effects. Clearly the integers
$n$ and $N$ correspond to the difference of radial and angular momentum quantum
numbers in Eq. (\ref{edi2}).

\section{Collective response (fixed surface)}

The zero-order response function (\ref{fsr}) gives only a first 
approximation to the nuclear response. When the residual interaction 
between nucleons is taken into account, a collective response function 
can be obtained by solving the Vlasov equation with appropriate 
boundary conditions \cite{bri,ads}. If the interaction is assumed to be 
of the octupole-octupole type,
\beq
\label{vres}
V({\bf r}_{1},{\bf r}_{2})=\kappa_{3} r_{1}^{3}\, r_{2}^{3}\, 
\sum_{M}Y^{*}_{3M}(\hat{\bf r}_{1})Y_{3M}(\hat{\bf r}_{2})
\eeq
the collective fixed-surface octupole response function is given by \cite{bdt}
\beq
\label{collfix}
{\cal R}_{3}(s)=\frac{{\cal R}_{3}^{0}(s)}{1-\kappa_{3} {\cal R}_{3}^{0}(s)}\,.
\eeq
 The parameter $\kappa_{3}$ specifies the strength of the residual 
 interaction, its value can be estimated in a self-consistent way, 
 giving (\cite{b&m}, p.557),
 \beq
 \kappa_{BM} =-\frac{4\pi}{3}\frac{m\omega_{0}^{2}}{A R^{4}}\approx 
 -1.\; 10^{-5} {\rm MeV/fm^{6}}
 \eeq
 for the octupole case. The parameter $\omega_{0}$ is  
 given by $\omega_{0}\approx 41 A^{-\frac{1}{3}}{\rm MeV}$ .
 Since this estimate is based on a harmonic oscillator mean field and 
 we are assuming a square-well potential instead, we expect some 
 differences. Hence we  shall determine the parameter $\kappa_{3}$
 phenomenologically, by requiring that the peak of the high-energy 
 octupole resonance agrees with experiment. This requirement 
 implies $\kappa_{3} \approx 2\kappa_{BM}$, which is in agreement with the 
 prescription obtained in the quadrupole case \cite{adm3}. In Fig.2 we report the 
 collective octupole strength function given by Eq. (\ref{collfix}), 
 with this value of $\kappa_{3}$. We can clearly see the effects of 
 collectivity that result in a shift and concentration of the 
 strength into two sharp peaks around
 20 Mev and 6-7 Mev. The experimentally observed 
 \cite{vdw}
 concentration of isoscalar octupole 
 strength in the two regions usually denoted by HEOR (high energy 
 octupole resonance) and LEOR (low energy octupole resonance) is 
 qualitatively reproduced, however the considerable strength 
 experimentally observed at lower energy (low-lying collective 
 states) is absent from our fixed-surface response function. Like for 
 the quadrupole response \cite{adm3}, we need to consider a different 
 solution of the linearized Vlasov equation (in which the nuclear 
 surface is allowed to vibrate \cite{ads}) in order to account for this
 feature of the response.
 
 The poles of the collective response function are 
 determined by the vanishing of the denominator in Eq. (\ref{collfix}).
 A solution of this equation at zero or purely imaginary frequency could be interpreted as 
 an instability of the spherical shape against an octupole-type 
 deformation of the ground state, however we can see that this could 
 happen only for interactions stronger than 
 \beq
 \kappa_{3cr}=-\frac{1}{{\cal R}_{3}^{0}(0)}\approx -3.14\, 10^{-5}{\rm 
 MeV/fm^{6}}\,.
 \eeq
 Our value of $\kappa_{3}$ is smaller (in absolute value) than this, so in 
 the present model the spherical shape is stable. This is valid for any 
 value of $A$, as long as we take $\kappa_{3}=2\kappa_{BM}$, since the product
 $\kappa_{BM}{\cal R}_{3}^{0}(0)$ is $A$-independent.
 
 The triangular orbits do not enter directly into the discussion of 
 this collective response, but the fact that they do not generate 
 pathologies at the level of the single-particle response ${\cal R}_{3}^{0}(s)$
 simplifies the discussion of the stability in the collective response 
 also.

\section{Collective response (moving surface)}

The collective fixed-surface response function (\ref{collfix}) does 
reproduce two important features of the octupole response: the HEOR 
and the LEOR, however it misses the experimentally observed low-lying 
collective $3^{-}$ states. In order to account for these low-lying 
states, a different solution of the linearized Vlasov equation has 
been proposed in Ref. \cite{ads}. In that approach the nuclear surface 
is allowed to vibrate according to the usual liquid-drop model 
expression 
\beq
\label{rot}
R(\theta,\varphi,t)=R+\sum_{LM}\delta R_{LM}(t) 
Y_{LM}(\theta,\varphi)\,.
\eeq
A self-consistency condition involving the surface tension is then 
used to determine the time-dependence of the additional collective 
variables $\delta R_{LM}$(t). Then, always for a separable residual 
interaction of the kind (\ref{vres}),
the collective fixed-surface response function (\ref{collfix})
is replaced by the following moving-surface response function 
\cite{adm3}

\beq
\label{rtil}
\tilde{{\cal R}}_{3}(s)=
{\cal R}_{3}(s)+ {\cal S}_{3}(s)\,,
\eeq
with ${\cal R}_{3}(s)$ still 
given by Eq. (\ref{collfix}), while ${\cal S}_{3}(s)$ gives the 
moving-surface contribution. For the separable interaction (\ref{vres})
the function ${\cal S}_{3}(s)$ can be evaluated explicitly as 
\cite{adm3}
\beq
\label{surfresp}
{\cal S}_{3}(s)=-\frac{R^{6}}{1-\kappa_{3} {\cal 
R}_{3}^{0}(s)}\:\: \frac{[\chi^{0}_{3}(s)+\kappa_{3}\varrho_{0}R^{3}
{\cal R}_{3}^{0}(s)]^{2}}{[C_{3}-\chi_{3}(s)][1-\kappa_{3} {\cal 
R}_{3}^{0}(s)]+\kappa_{3}R^{6}[\chi^{0}_{3}(s)+\varrho_{0}R^{3}]^{2}}\,,
\eeq
with $C_{3}= 10\sigma R^{2}+(C_{3})_{coul}$ ($\sigma\approx 1 {\rm MeV\,fm^{-2}}$
is the surface tension parameter obtained from the mass formula, 
$(C_{3})_{coul}$ gives the Coulomb contribution to the restoring force) and 
$\varrho_{0}=A/\frac{4\pi}{3}R^{3}$ the equilibrium density.

The functions $\chi^{0}_{3}(s)$ and $\chi_{3}(s)$ are defined as 
\beq
\label{chiok}
\chi^{0}_{3}(s) =\frac{9A}{4\pi}\sum_{n N}
C_{3N}^{2}\int_{0}^{1} dx x^{2}~
s_{nN}(x){(-)^{n} ({Q^{(3)}_{nN}(x)/R^{3})}
\over {s+i\varepsilon-s_{nN}(x)}},
\eeq
and
\beq
\label{chik}
\chi_{3}(s)=-\frac{9A}{2\pi}\epsilon_{F}\,(s+i\varepsilon)
\sum_{n N} C_{3N}^{2} \int_{0}^{1} dx x^{2}~ \frac{1}
{s+i\varepsilon-s_{nN}(x)}\,,
\eeq
their structure is similar to that of the zero-order propagator 
(\ref{fsr}).

In Fig.3 we report the octupole strength function given by the 
moving-surface response function (\ref{rtil}) and compare it to the 
collective fixed-surface response given by Eq. (\ref{collfix}). The 
most relevant change induced by the moving surface is the large 
double hump appearing at low energy. This feature is in qualitative 
agreement both with experiment \cite{vdw} and with the result 
of RPA-type calculations \cite{lb,ll}, moreover is rather similar to 
that found in Ref. \cite{adks} within the same model, but with  
different excitation operator and residual interaction. We interpret 
this low-energy double hump as a superposition of surface vibrations 
and LEOR. In order to explain why we do not obtain one or more sharp $3^{-}$ 
states at low energy, we have to analyze our moving-surface response 
function in some detail. Since the explicit expression 
(\ref{surfresp}) looks rather involved, we consider the limit of 
non-interacting nucleons. If we let $\kappa_{3}\to 0$,
the function (\ref{surfresp}) becomes
\beq
\label{zsurf}
{\cal 
S}^{0}_{3}(s)=-R^{6}\frac{[\chi^{0}_{3}(s)]^{2}}{C_{3}-\chi_{3}(s)}\,,
\eeq
and the full moving-surface response function (\ref{rtil})
\beq
\label{rtilz}
\tilde{{\cal R}}^{0}_{3}(s)=
{\cal R}^{0}_{3}(s)+ {\cal S}^{0}_{3}(s)\,.
\eeq

In Fig.4 we show the octupole response function evaluated both for 
$\kappa_{3}=2\kappa_{BM}$ (solid curve) and for $\kappa_{3}=0$ 
(dashed curve). We can see from this figure that, while the residual 
interaction changes drastically the response in the giant resonance 
region, at low energy the 
octupole response is affected only very slightly by the  
interaction (\ref{vres}). Thus the low-energy octupole  response can be 
analyzed by using the simpler formula (\ref{zsurf}), rather than Eq. 
(\ref{surfresp}).

The eigenfrequencies of low-energy collective modes are approximately 
determined by the vanishing of the denominator in Eq. (\ref{zsurf}), 
moreover at low frequency the function $\chi_{3}(\omega)$ can be 
expanded as \cite{adm1}
\beq
\label{lfe}
\chi_{3}(\omega)=i\omega\gamma_{3}+D_{3}\omega^{2}+\ldots \,,
\eeq
implying that the parameters $\delta R_{3M}(t)$, that describe octupole surface
vibrations in Eq. (\ref{rot}), approximately satisfy an equation of 
motion of the damped oscillator kind:
\beq
\label{odos}
D_{3}\delta \ddot{R}_{3M}(t) +\gamma_{3} \delta 
\dot{R}_{3M}(t)+C_{3}\delta R_{3M}(t)=0\,.
\eeq

It has already been pointed out in \cite{adm1} that for $A=208$ the 
numerical values of parameters in this equation are such that this 
oscillator is actually overdamped, moreover, always in
\cite{adm1}, the coefficients $\gamma_{L}$ have been evaluated 
analyticallty in the low-frequency limit, giving (for a generic $L$)
\beq
\label{gamma}
\gamma_L =\gamma_{wf}~2{(4\pi)^2\over{2L+1}}\sum_{N=1}^L {1\over N}
\mid Y_{LN}({\pi\over 2},{\pi\over 2})\mid^2
\sum_{n=1}^{+\infty} \cos\alpha_{nN}\sin^3 \alpha_{nN}
\Theta({\pi\over 2}-\alpha_{nN})\,,
\eeq
with $\gamma_{wf}=\frac{3}{4}\varrho_{0}p_{F}R^{4}$ and 
$\alpha_{nN}=\frac{n}{N}\pi$. The angles $\alpha_{nN}$ are related to the nucleon 
trajectories as discussed in Sect. II. In our case the 
coefficient $\gamma_{L=3}$ gets a contribution only from the term with $n=1$ and 
$N=3$, thus we see that only nucleons moving along closed triangular 
trajectories can contribute to the damping of octupole 
surface vibrations. In order to check that indeed the closed 
triangular orbits are the main source of Landau damping in the 
low-energy octupole response, we study also the response that is obtained when 
the contribution of these orbits is excluded. This exclusion can be 
made simply by avoiding a small interval of length $2\Delta$ about the 
value $x=\frac{\sqrt{3}}{2}$ in the $x$-integration in Eqs. 
(\ref{fsr}), (\ref{chiok}) and(\ref{chik}). In Fig.5 we show the 
response obtained with a value of $\Delta=0.02$. The shape of the 
low-energy hunp changes dramatically because of the lack of damping 
due to the missing triangular trajectories. A very sharp peak is now 
developed at low energy. The position of this peak is influenced by 
the value of the mass parameter $D_{3}$, omitting the triangular 
orbits changes also the numerical value of this parameter.
Once again we see 
that the closed triangular orbits, rather than generating a shape 
instability in the octupole channel, are the main source of Landau 
damping in this response function. Of course quantum effects can play 
an essential role at this level since in quantum mechanics the  
angular momentum is quantized and there could be no value of angular 
momentum corresponding to triangular trajectories.

 We would also like to point out that our moving-surface response 
function can display a shape instability in the octupole channel, 
which can arise if the restoring force parameter $C_{3}$ vanishes. This could 
happen if the repulsive Coulomb term $(C_{3})_{coul}$ exactly balances the 
attractive surface-tension part of the restoring force. In this case 
our moving-surface response function developes a pole at the origin, 
as shown in Fig.6. For odd multipolarities the instability condition 
in our model is the same as for the liquid-drop model \cite{adm1}.

Before concluding, in Fig.7 we compare our moving surface response 
function with the response function of the overdamped oscillator 
(\ref{odos}). This figure supports our interpretation of the 
lower-energy
hump as due to overdamped surface vibrations.

\section{Conclusions}

We have analyzed the octupole response function of a hypothetical 
heavy nucleus containing $A=208$ nucleons in a square-well potential 
by employing a semiclassical model that relates the response to 
features of the classical trajectories of nucleons. Triangular 
trajectories are particularly relevant for the octupole response 
because, at the single-particle level, a simple combination of their 
characteristic frequencies, appearing in this response function, can vanish.
This fact had already been 
noticed long ago and some speculations had been made on the possible 
connection between these classical orbits and the onset of a shape 
instability in heavy spherical nuclei against octupole-type 
deformations \cite{b&m}. Our detailed calculations performed in a 
semiclassical approach show that the vanishing of the 
eigenfrequencies associated with triangular trajectories has no 
consequences on the stability of the spherical shape in heavy nuclei, 
rather, triangular orbits are essential in providing a damping of low-energy
octupole excitations.

\begin{figure} Fig.1 Semiclassical octupole strength function analogous to 
quantum single-particle strength function. Note the strength 
concentration in the $1\hbar\omega$ and $3\hbar\omega$ regions. 
Calculations are for $A=208$ nucleons in a square well potential of 
radius $R=1.2 A^{\frac{1}{3}}\;{\rm fm}$.
\end{figure}

\begin{figure} Fig.2 Collective strength function (solid curve) evaluated 
for fixed-surface boundary conditions. The residual interaction 
parameter is $\kappa_{3}=-2.\; 10^{-5} {\rm MeV/fm^{6}}$. The dashed curve is the 
strength function of Fig.1.
\end{figure}

\begin{figure} Fig.3 The solid curve shows the octupole strength 
function evaluated by using moving-surface boundary conditions in the 
Vlasov equation, the dashed curve instead, corresponds to 
fixed-surface boundary conditions.
\end{figure}

\begin{figure} Fig.4  Moving-surface strength function for
$\kappa_{3}=-2.\; 10^{-5} {\rm MeV/fm^{6}}$ (solid) and for $\kappa_{3}=0$ (dashed).
\end{figure}

\begin{figure} Fig.5 Moving-surface strength function with (solid) 
and without (dashed) contribution of closed triangular orbits. 
\end{figure}

\begin{figure} Fig.6 Collective moving-surface strength function 
(solid) dispaying a divergence at vanishing excitation energy when 
$C_{3}\to 0$ (dashed). This divergence is interpreted as a shape 
instability.
\end{figure}

\begin{figure} Fig.7 Collective moving-surface strength function 
(solid) compared to overdamped oscillator strength function (dashed) 
with appropriate parameters. The two functions practically coincide 
for $E<2\;{\rm MeV}$.
\end{figure}
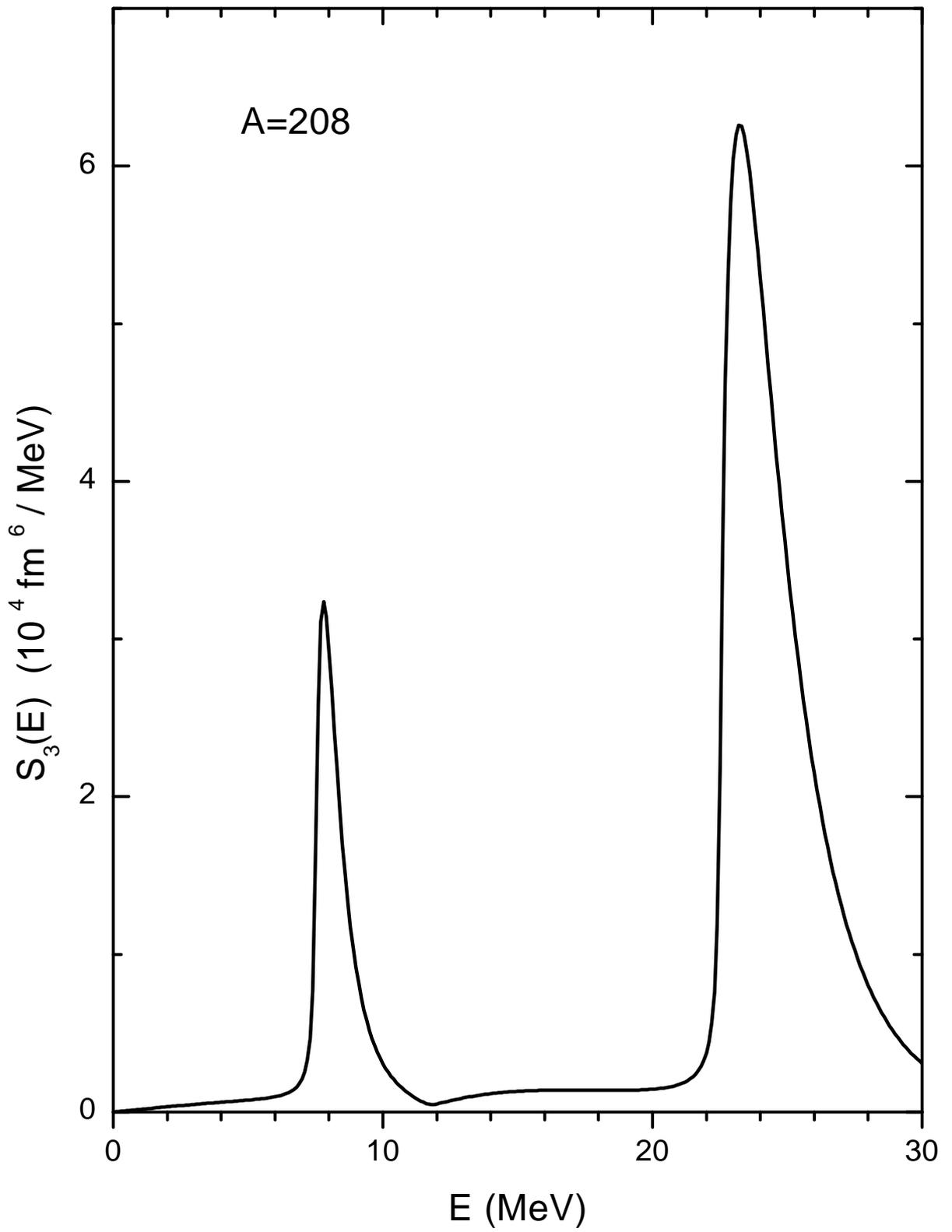
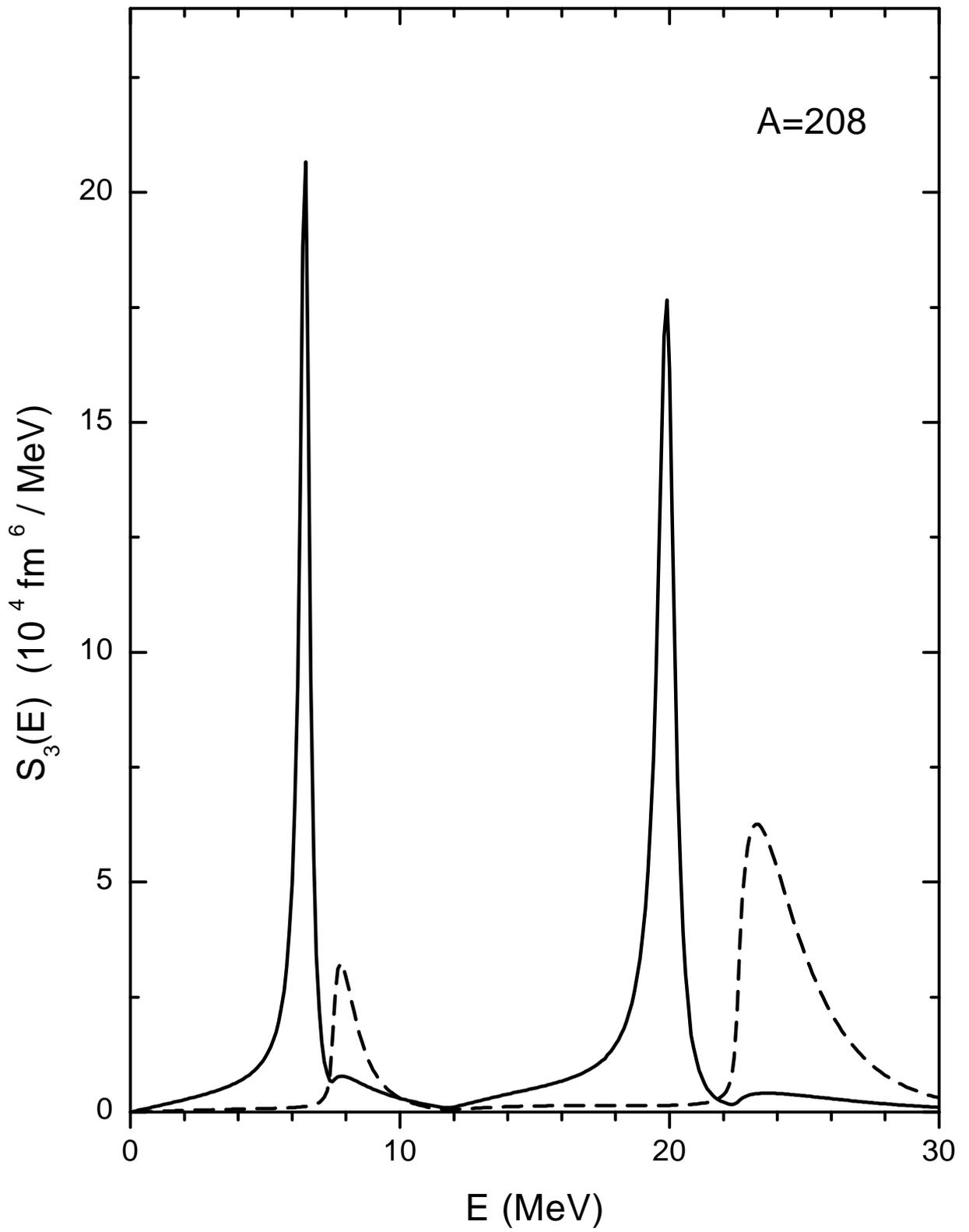
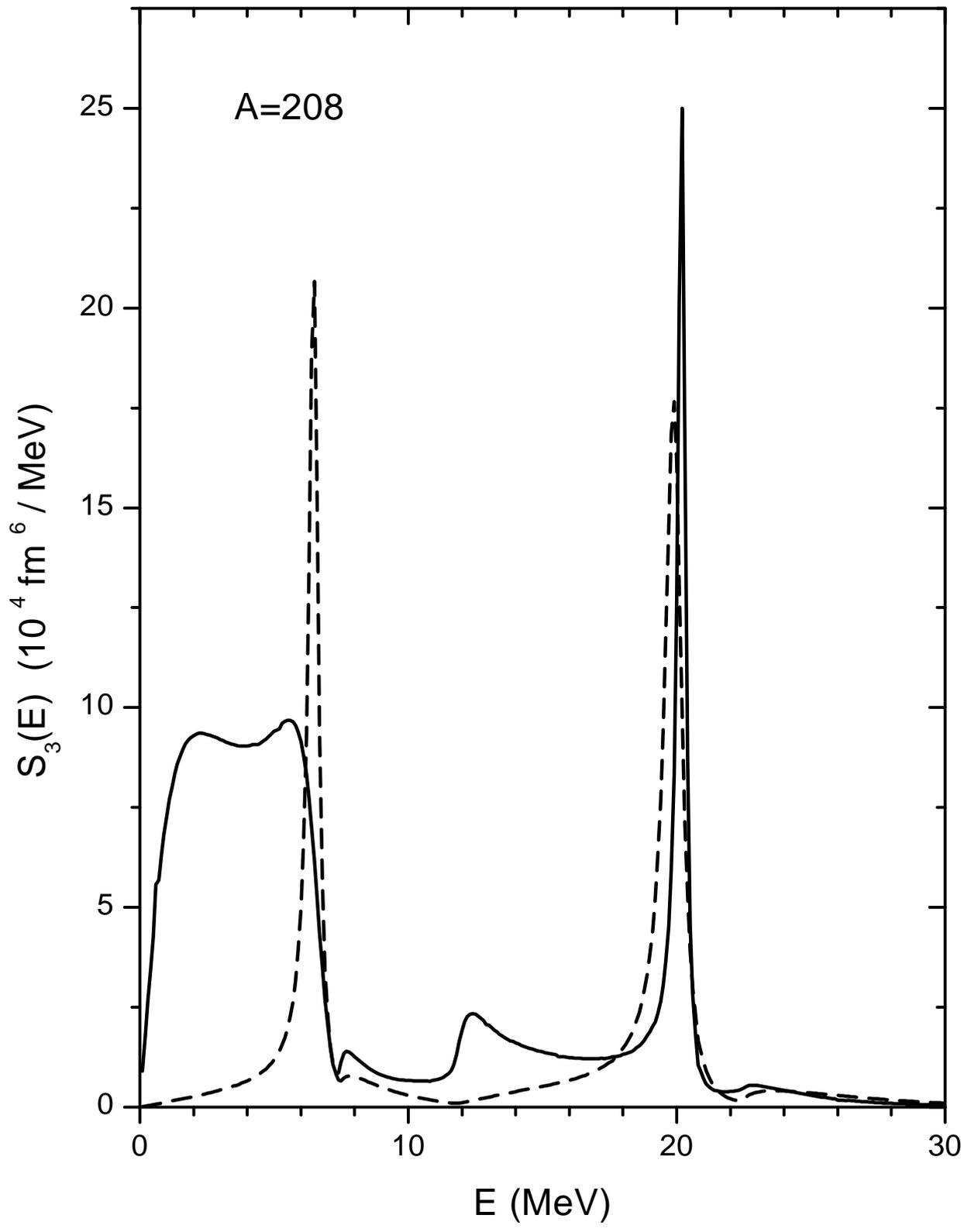
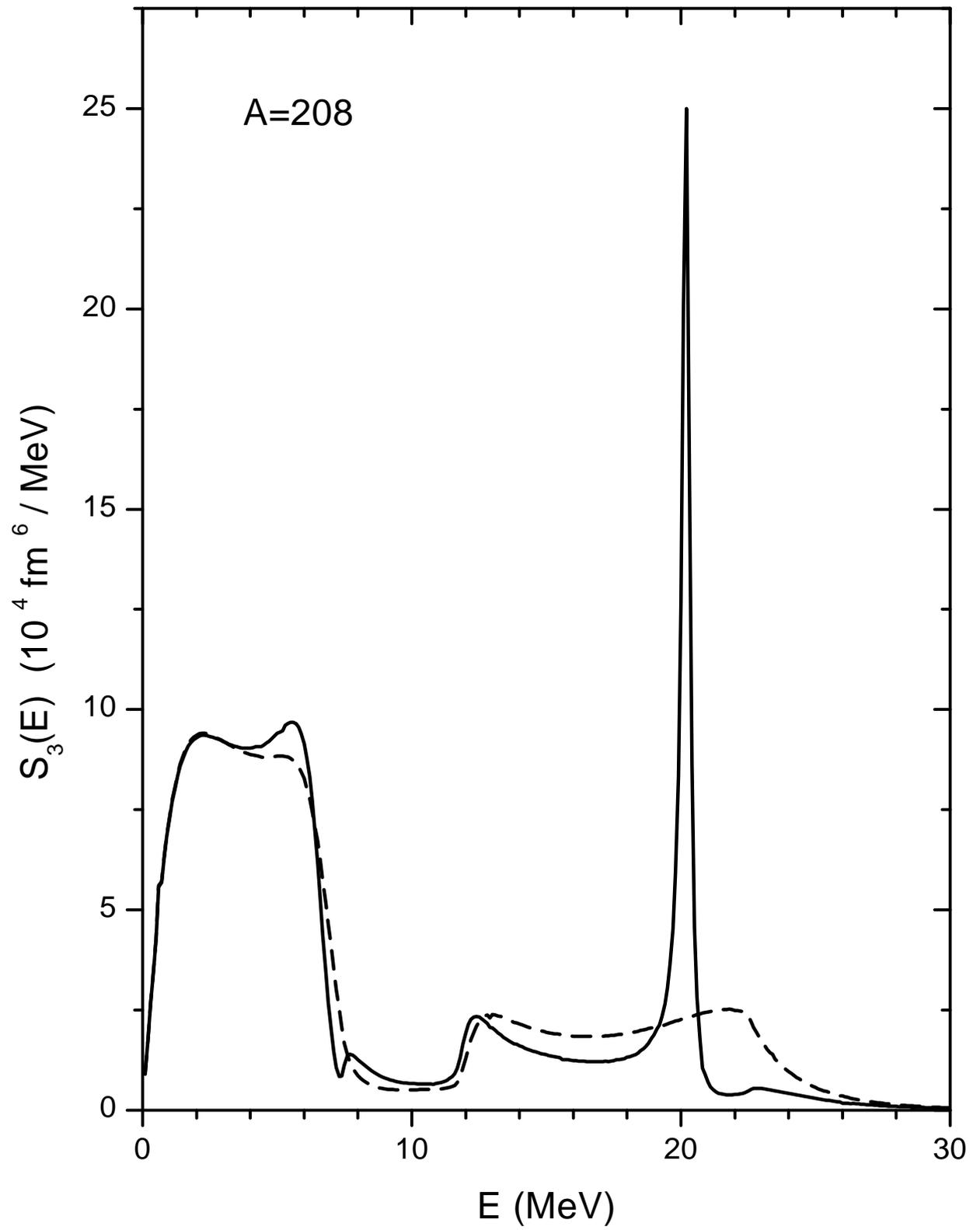
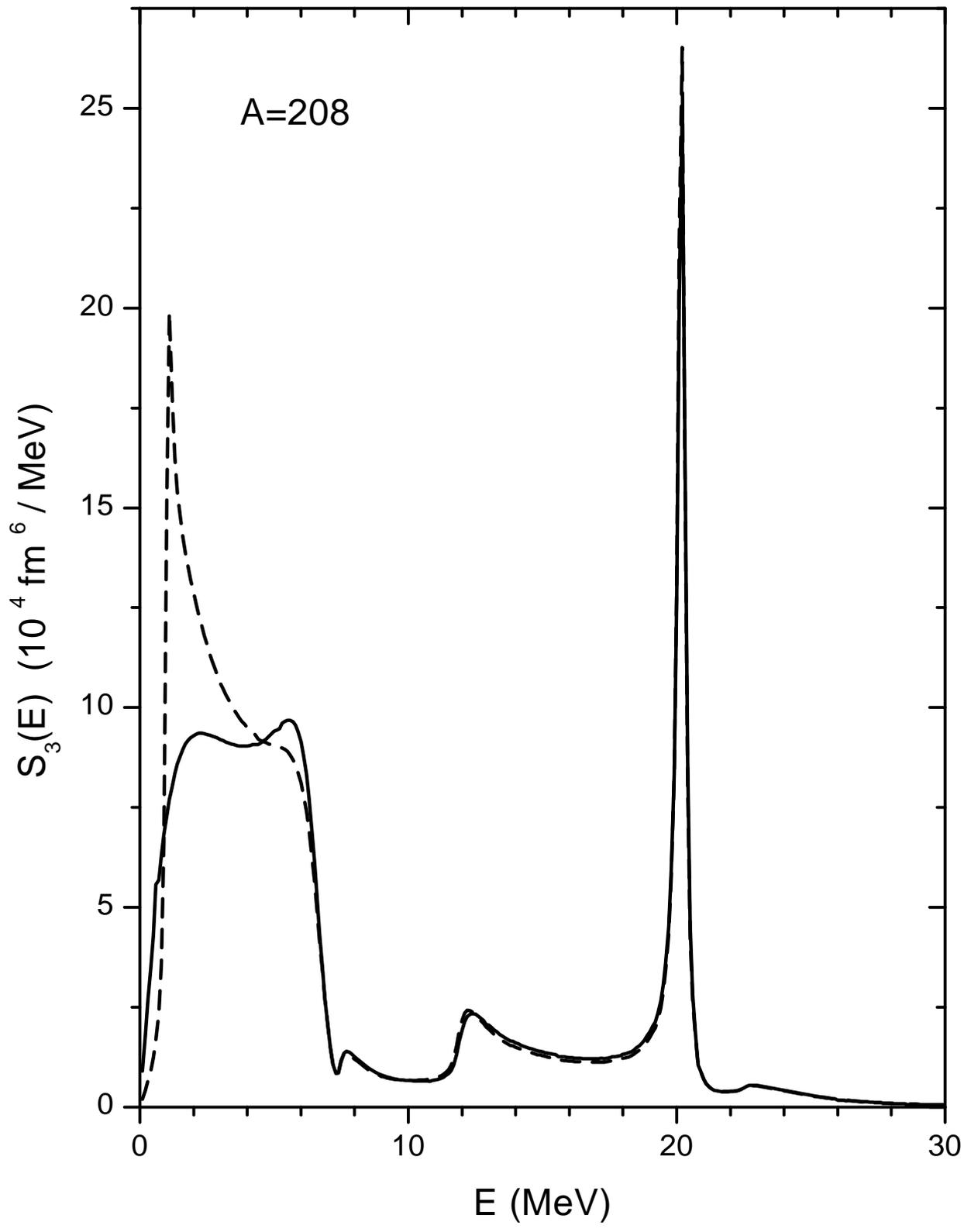
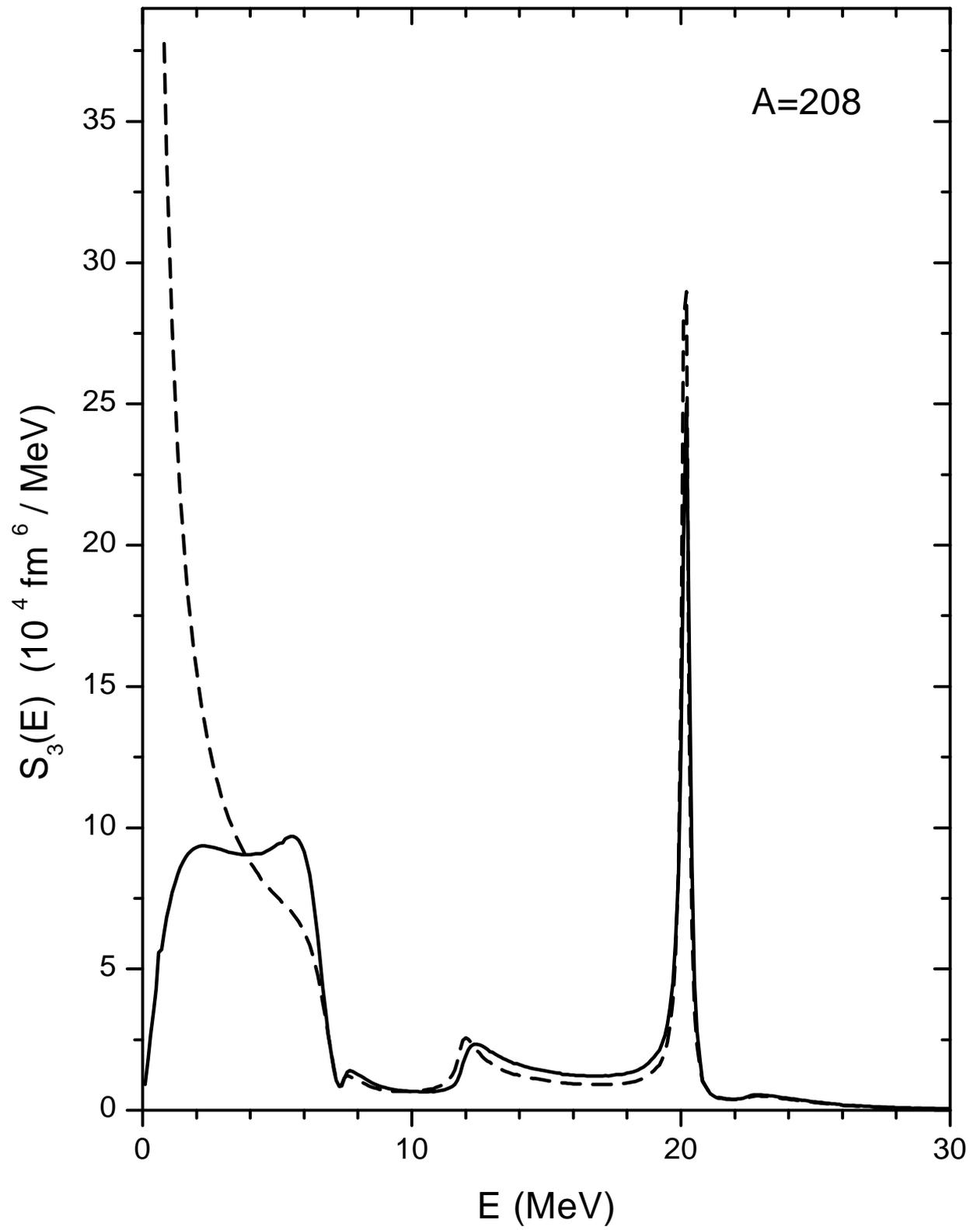
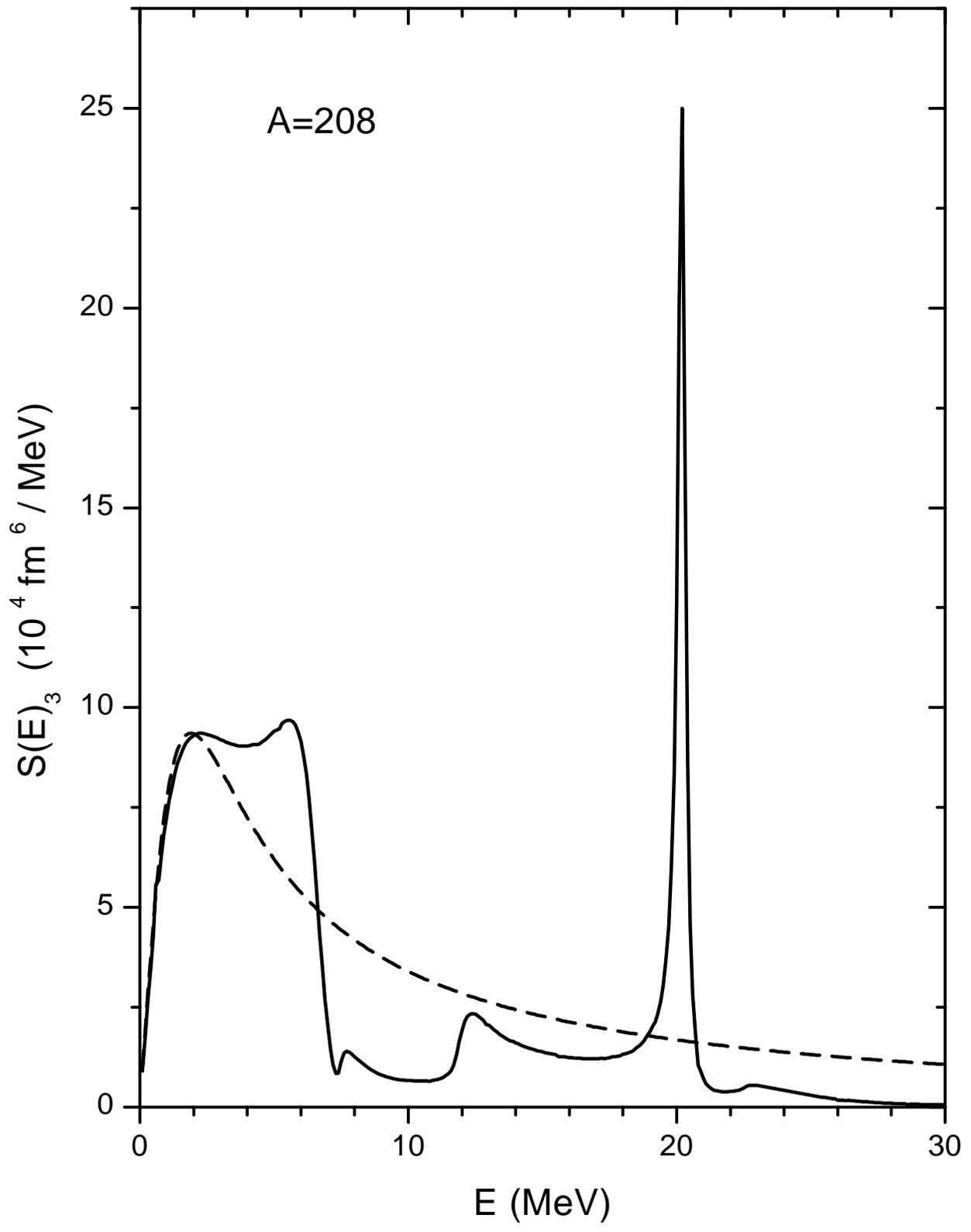
\end{document}